\documentclass{article}

\usepackage{arxiv}

\usepackage[utf8]{inputenc} 
\usepackage[T1]{fontenc}    
\usepackage{hyperref}       
\usepackage{url}            
\usepackage{booktabs}       
\usepackage{amsfonts}       
\usepackage{nicefrac}       
\usepackage{microtype}      
\usepackage{lipsum}		
\usepackage{graphicx}
\usepackage[square,sort,comma,numbers]{natbib}
\usepackage{doi}

\title{Modeling and Predicting Blood Flow Characteristics through Double Stenosed Artery from CFD simulation using Deep Learning Models}

\author{{Ishat Raihan Jamil} \\
	Department of Mechanical Engineering\\ Bangladesh University of Engineering and Technology\\ 
    Dhaka, Bangladesh\\
	\texttt{ishatraihanjamil@gmail.com} \\
	\And
	{Mayeesha Humaira} \\
	Department of Computer Science and Engineering\\ 
    Ahsanullah University of Science and Technology\\ 
    Dhaka, Bangladesh\\
	\texttt{mayeeshahumaira@gmail.com} \\
}

\renewcommand{\shorttitle}{\textit{arXiv} Template}

\hypersetup{
pdftitle={A template for the arxiv style},
pdfsubject={q-bio.NC, q-bio.QM},
pdfauthor={David S.~Hippocampus, Elias D.~Striatum},
pdfkeywords={First keyword, Second keyword, More},
}

\begin{document}
\maketitle

\begin{abstract}
	Establishing patient-specific finite element analysis (FEA) models for computational fluid dynamics (CFD) of double stenosed artery models involves time and effort, restricting physicians' ability to respond quickly in time-critical medical applications. Such issues might be addressed by training deep learning (DL) models to learn and predict blood flow characteristics using a dataset generated by CFD simulations of simplified double stenosed artery models with different configurations. When blood flow patterns are compared through an actual double stenosed artery model, derived from IVUS imaging, it is revealed that the sinusoidal approximation of stenosed neck geometry, which has been widely used in previous research works, fails to effectively represent the effects of a real constriction. As a result, a novel geometric representation of the constricted neck is proposed which, in terms of a generalized simplified model, outperforms the former assumption. The sequential change in artery lumen diameter and flow parameters along the length of the vessel presented opportunities for the use of LSTM and GRU DL models. However, with the small dataset of short lengths of doubly constricted blood arteries, the basic neural network model outperforms the specialized RNNs for most flow properties. LSTM, on the other hand, performs better for predicting flow properties with large fluctuations, such as varying blood pressure over the length of the vessels. Despite having good overall accuracies in training and testing across all the properties for the vessels in the dataset, the GRU model underperforms for an individual vessel flow prediction in all cases. The results also point to the need of individually optimized hyperparameters for each property in any model rather than aiming to achieve overall good performance across all outputs with a single set of hyperparameters.

\end{abstract}


\section{Introduction}
Cardiovascular diseases (CVDs) are the most common causes of death around the world. Heart attacks are typically sudden occurrences caused by the narrowing and blockage of blood vessels\footnote{https://www.who.int/news-room/fact-sheets/detail/cardiovascular-diseases-(cvds)}. A stenosed artery refers to the narrowing of a blood vessel caused by the deposition of atherosclerotic plaque on the inner walls of the arterial lumen \cite{BM7_2, BM7_3}. These cholesterol and fatty deposits lead to a swollen and inflamed inner arterial wall which restricts the flow of oxygenated blood cells, nutrients, and other essential substances from reaching the heart muscles \cite{BM7_4, BM7_5}. Cholesterols have been shown to accelerate the formation of plaque in arteries, eventually obstructing the bloodstream and altering hemodynamics \cite{BM7_9}. When the plaque ruptures, the accumulated fatty acids, platelets, and dead cells may coagulate, resulting in thrombosis formation \cite{BM7_10}. In the case of a coronary or cerebral artery, the blood clot may have fatal consequences since it will also cut off the blood flow to the cerebral region of the brain or the myocardial heart wall \cite{BM7_11}. The likelihood of developing thrombosis is highly dependent on the thickness of the plaque, the characteristics of infected blood, and blood pressure \cite{BM7_12}.

Numerous studies observed pulsing flow behavior and constant dampening of its oscillations, which they attributed to the flexibility of blood vessels \cite{BM7_17}. Coronary artery disorder is critical in hemodynamics because it alters the flow pattern, resulting in variations in the wall pressure and shear stress of the arteries. As a result, health researchers must determine the flow velocity and amount of shear stress in arteries. A substantial part of the published research \cite{BM7_18} focused on the physiological origins of the disease as they relate to blood vessels. However, few have made strides in understanding the underlying physics of the illness in order to better understand the cause and, as a result, paving the way to less invasive and more long-term treatments. Medical imaging can be utilized to visualize the areas of fatty deposits inside artery walls, but it is not capable of providing numerical data in the same way that computational fluid dynamics (CFD) simulations are capable of providing. According to Kompatsiaris et al. \cite{BP4_1} and Liu et al. \cite{BP4_2}, computational simulations can offer an in-depth evaluation of flow resistance owing to wall shear stress ($WSS$) on blood vessel walls, blood flow rates, and pressure changes. CFD results generated by modeling vessels in the relevant regions may be compared to the reliability of mathematical data. It is possible to develop a less invasive dependable method for medical diagnosis by integrating physician expertise with data derived from realistic computational fluid dynamics models. Owing to the vessel's small dimensions, in vitro, and in vivo flow field experiments are not representative and accurate. Thus, with improved software development and computer efficiency, CFD may replace such experimental approaches. CFD has been already used in several studies involving blood flow through vessels. Fazlay et al. \cite{ABM2} used CFD to show that following a double stenosed region in an artery, blood flow is hampered significantly at maximum systolic velocity and acceleration. Jianhuang et al. \cite{BP4} coupled transient blood flow with elastic artery to evaluate unsteady flow characteristics along its length using computational fluid dynamics. Mukesh et al. \cite{BM6} used an in-house CFD solver to verify and simulate blood flow via a stenosed artery. Mehdi et al. \cite{BP7} compared several turbulent models from blood flow through vessels and concluded that inaccuracies are introduced by assuming the flow to be laminar.

Setting up patient-specific finite element analysis (FEA) models for CFD takes time and effort, limiting quick response to physicians in time-sensitive medical applications. As such, Liang et al. \cite{Deep_WSS} created deep learning (DL) algorithm to predict aortic stress distributions. The DL model was developed to use FEA data and directly produce aortic wall stress distributions, skipping the FEA calculation step entirely. Such progressions in computer science open up new horizons for development. For instance, if the diameters of the stenosed aortic vessel are considered at regular intervals, a special kind of artificial neural network called recurrent neural network (RNN) can be implemented. RNN’s internal memory allows them to comprehend sequential data. Their ability to retain crucial details about the preceding step, such as aortic diameter, enables them to predict occurrences in the next step such as $WSS$, blood pressure or flow velocity accurately. Previously RNN has been successfully implemented for malware classification \cite{RNN1}, 3D shape generation \cite{RNN2}, traffic forecasting \cite{RNN3}, and speech enhancement \cite{RNN4}.

\section{Computational Fluid Dynamics Simulations}
Several data are necessary to construct a dataset in order to implement artificial intelligence (Ai). The lack of sufficient medical data on blood flow patterns in doubly stenosed arteries necessitates the use of computational approaches to generate the data. This allows for the exploration and visualization of the variations in blood flow behavior induced by several combinations of stenosis at varying distances apart. This section describes the processes used to set up CFD simulations and compares a couple of simplified models to identify which one best depicts the real flow characteristics, using an actual double stenosed artery model as a reference.

\subsection{Governing Equations}
If the Navier–Stokes equation is interpreted as the sum of an average and an oscillating component for each variable, then the continuity and Reynolds averaged Navier–Stokes equations (RANS) are as follows:
\begin{equation}
   \frac{\partial \bar u_i}{\partial x_i} = 0\label{eq1}
\end{equation}
\begin{equation}
   \frac{D\bar u_i}{Dt} = -\frac{1}{\rho}\frac{\partial \bar P}{\partial x_i} + \frac{\partial}{\partial x_j}\left(\frac{\mu + \mu_T}{\rho}\left(\frac{\partial \bar u_i}{\partial x_j} + \frac{\partial \bar u_j}{\partial x_i}\right)\right)\label{eq2}
\end{equation}
where $\mu$ represents the summation of laminar $\mu_0$ and turbulent $\mu_T$ viscosities:
\begin{equation}
   \mu = \mu_0 + \mu_T \label{eq3}
\end{equation}
For $K$-$\varepsilon$ standard turbulence model, $\mu_T$ is computed as:
\begin{equation}
   \mu_T = \rho c_\rho \frac{K^2}{\varepsilon} \label{eq4}
\end{equation}
where $K$ represents the turbulence kinetic energy and $\varepsilon$ denotes the rate of turbulence dissipation.

A straight double stenosed artery can be simply modeled as a tube with a diameter $D$ with stenosed necks $S_1$ and $S_2$ separated by a distance of $L$. The degree of obstruction of the stenosed regions can be expressed as follows:
\begin{equation}
    \%S= \frac{D - d}{D} \times 100\% \label{eq5}
\end{equation}
where $d$ is the lumen diameters at neck $S$. The fraction of lumen opening at the neck then can be addressed as:
\begin{equation}
    Fraction\ of\ lumen\ opening = 1 - \%S/100 \label{eq6}
\end{equation}

\subsection{Simulation Setup}
Solidworks is used to model the arteries for this study. The blood vessels are assumed to be rigid with the no-slip condition at the arterial wall. Ansys Fluent software is used to set up the simulation and solve the RANS equations utilizing the finite volume method (FVM). The second-order upwind scheme was employed to spatially discretize the governing equations and the SIMPLE method was used to manage the pressure-velocity decoupling \cite{BP6}. Blood is considered to be an incompressible fluid with a density of $1050\ kg/m^3$ and a viscosity of $0.0033\ Pa.s$ \cite{BP7}. The inlet flow velocities are obtained from the velocity profile presented by Fazlay et al. \cite{ABM2}. The authors pointed out five particular velocities from the waveform: $0.21\ m/s$, $0.33\ m/s$, $0.28\ m/s$, $0.14\ m/s$, and $0.09\ m/s$ at maximum systolic acceleration, systolic velocity, systolic deceleration, diastolic velocity, and at minimum systolic velocity respectively. Due to the presence of plasma, platelets, and suspended cells, blood has the characteristics of non-Newtonian fluid \cite{BP6}. However, numerous previous CFD research treated blood as a Newtonian fluid \cite{ABM1, ABM2, BP7}. In fact, Ku et al. \cite{ABM1 16} observed that for Reynolds numbers ($Re$) ranging from $110$ to $850$ in big arteries, the non-Newtonian impact of blood is insignificant. As such, blood is deemed Newtonian in this analysis since $Re$ stays within this range. The simulations are carried out with a time step of $0.0001\ s$ and a mesh element size of $0.112\ mm$.

\begin{figure}[h!]
	\centering
	\includegraphics[width=0.5\linewidth]{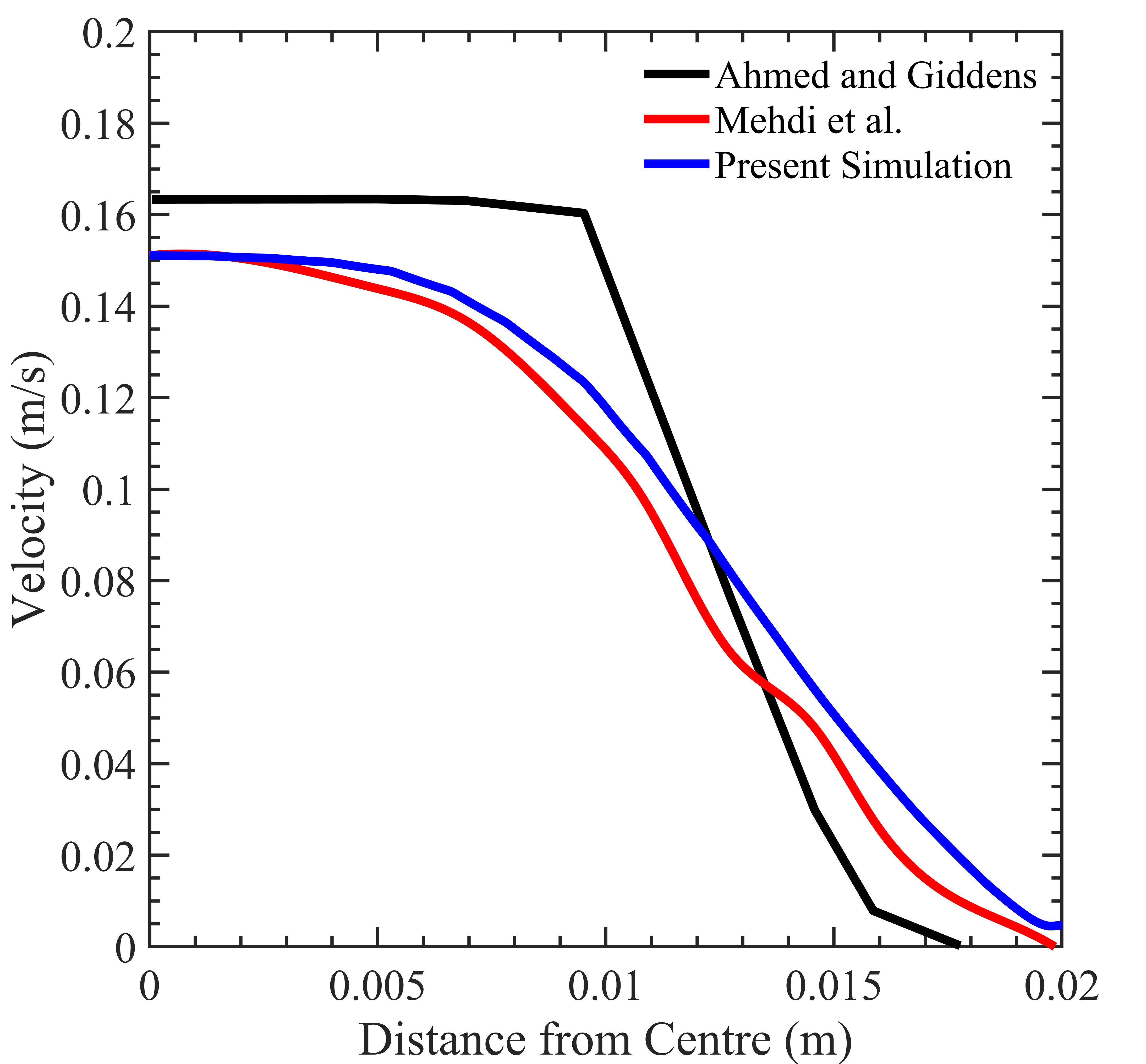}
	\caption{Validation test comparing velocity profile of present simulation with simulation result of Mehdi et al. and experimental results of Ahmed and Giddens at a distance equal to the vessel's diameter D away from the stenosed neck.}
	\label{fig1}
\end{figure}

Fig. \ref{fig1} shows the velocity profile for the sinusoidal stenosed artery model presented by Fazlay et al. \cite{ABM2} at a distance equal to the vessel's diameter $D$ away from the stenosed neck using the aforementioned blood flow characteristics and the $K$-$\varepsilon$ standard turbulence model. The velocity curve closely matches the velocity profile of the model suggested by Mehdi et al. \cite{BP7} and has a good agreement with the experimental results of Ahmed and Giddens \cite{BP7_15}. As a result, further simulations are performed using this particular turbulence model.

\subsection{Modeling Double Stenosed Artery}
\begin{figure}[h!]
	\centering
	\includegraphics[width=0.6\linewidth]{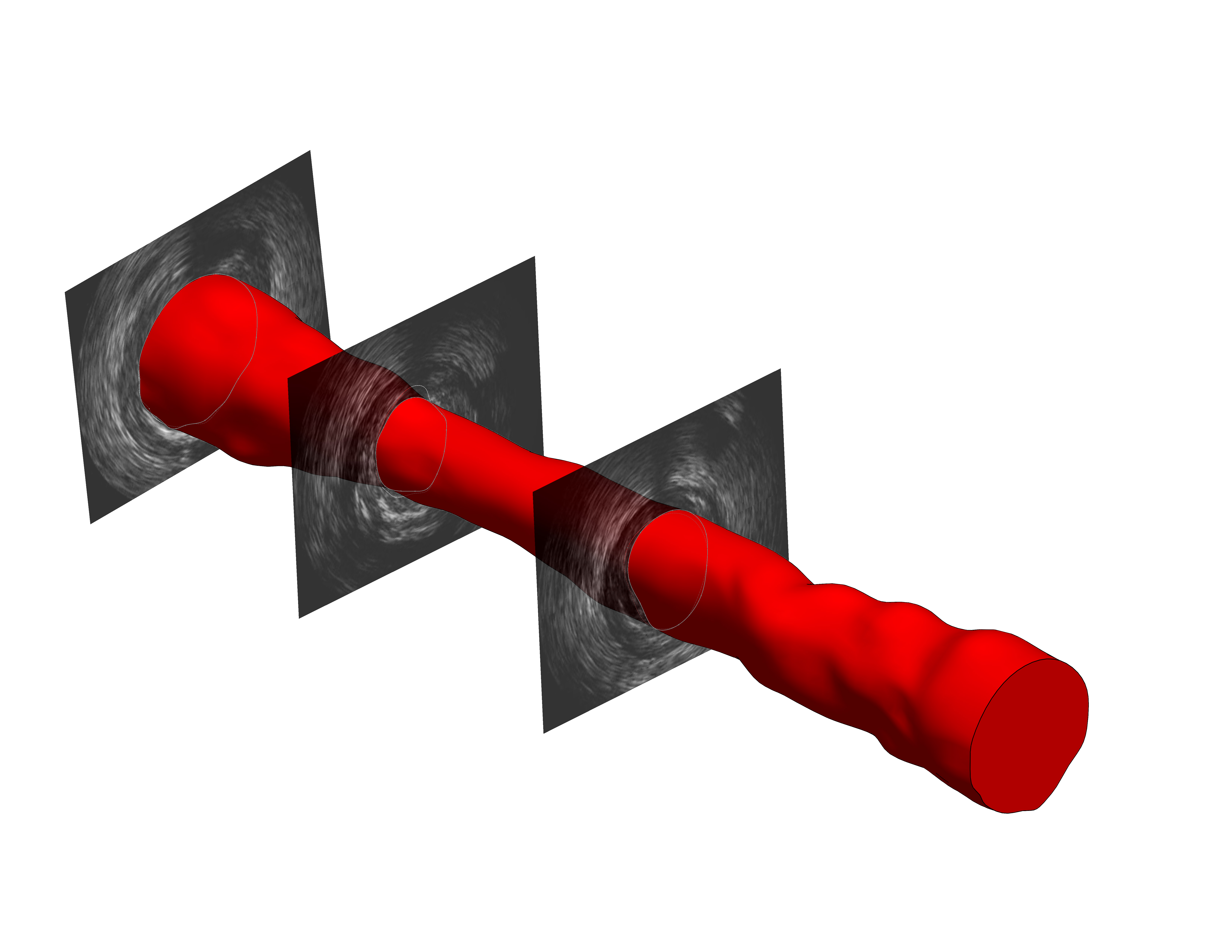}
	\caption{Modeling of an actual double stenosed artery from IVUS images. For clarity, only a few IVUS images are shown.}
	\label{fig2}
\end{figure}

\begin{figure}[h!]
	\centering
	\includegraphics[width=0.6\linewidth]{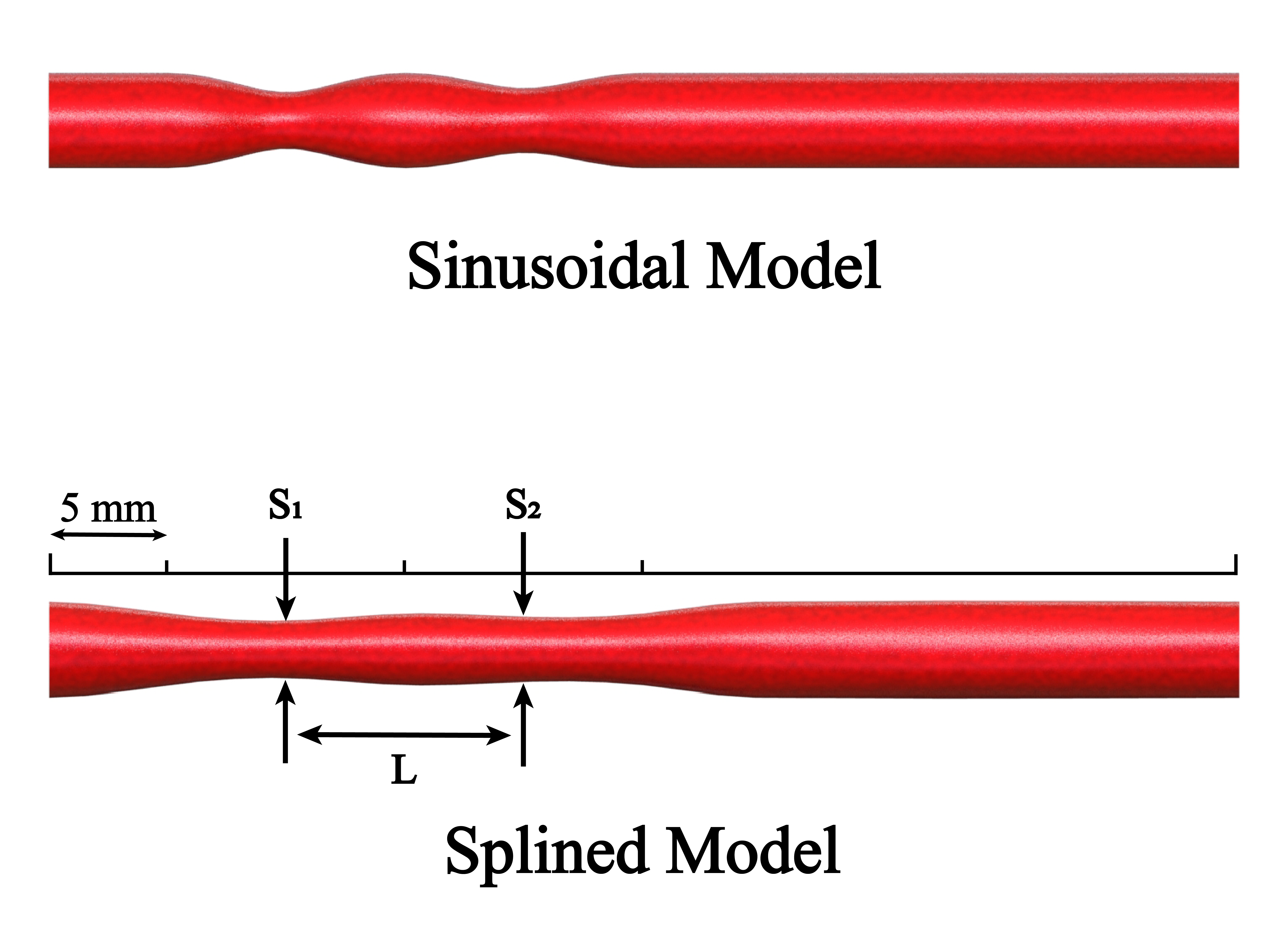}
	\caption{Comparison between the sinusoidal and splined representation of stenosed neck geometry in a simplified model of the actual double stenosed artery.}
	\label{fig3}
\end{figure}

\begin{figure}[h!]
	\centering
	\includegraphics[width=0.98\linewidth]{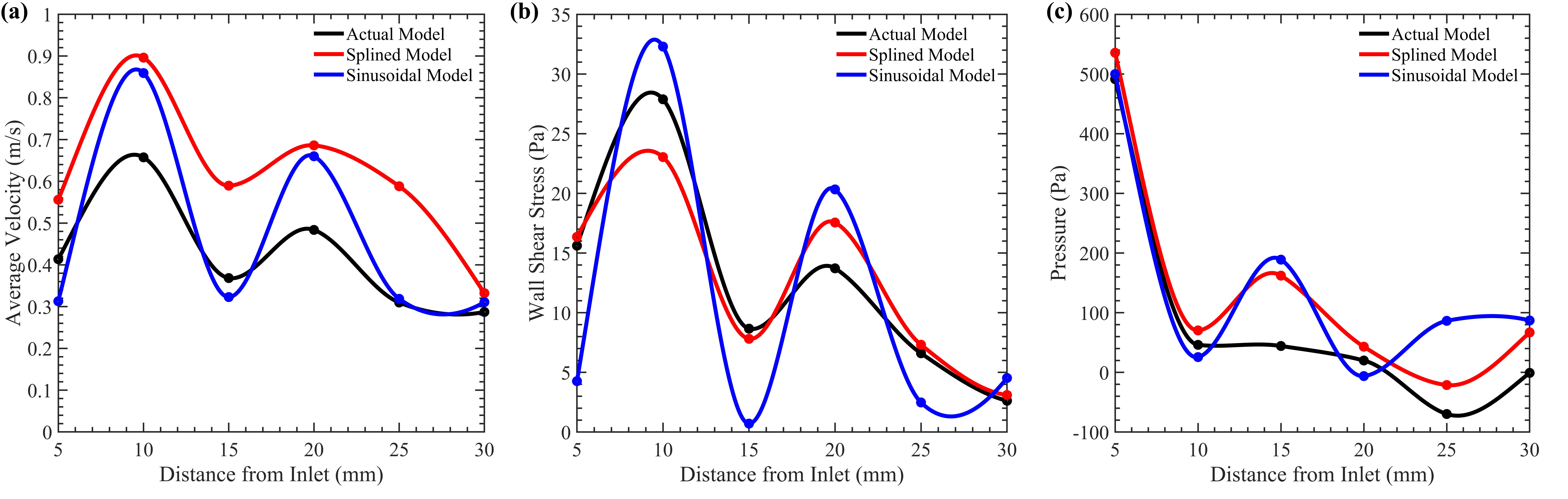}
	\caption{Comparison of flow characteristics through the actual and simplified double stenosed artery models.}
	\label{fig4}
\end{figure}

Fig. \ref{fig2} illustrates the modeling of a straight section of an actual doubled stenosed artery obtained via intravascular ultrasound (IVUS) imaging using a $3\ fr$ catheter and a $1\ mm$ guidewire, with a pullback rate of $1\ mm/s$. It has an average non-stenosed hydraulic diameter of approximately $4\ mm$, with $40.25\%$ and $32\%$ stenosis situated $10\ mm$ apart. Fig. \ref{fig3} shows two simplified representations of the actual artery with similar stenoses. Unlike the sinusoidal equation-generated model \cite{ABM2} for a similar configuration, the actual model exhibits a gradual decrease in lumen diameter, as visible from fig. \ref{fig4}, while fig. 4 illustrates the variation in blood flow patterns through them. This simplification has a significant effect on the flow characteristics of blood. As demonstrated by fig. \ref{fig4}, the steep sinusoidal stenosed edges exhibit a wide variation in average velocity ($V_{avg}$), wall shear stress ($WSS$), and pressure from those obtained from the actual model simulation for input velocity of $0.3\ m/s$. As such, a more representative model is required. Fig. \ref{fig3} presents another model denoted as the splined model. This model features a $25\%$ stenosis region $5\ mm$ upstream and $5\ mm$ downstream of the main stenosed neck. When such circular cross-sections are joined with spline guidelines, the simplified model captures the actual model’s naturally formed gradual stenosis characteristics. This results in a $WSS$ curve that is more similar to that of the actual artery model as can be seen in fig. \ref{fig4}. The $V_{avg}$ curve also shows a similar trend but is visibly higher due to the absence of surface irregularities to retard the flow as in the actual model. On the other hand, both simplified models fail to represent the actual model's pressure fluctuations effectively, especially at the diverging sections following the stenosed necks. 

\begin{figure*}[h!]
	\centering
	\includegraphics[width=0.95\linewidth]{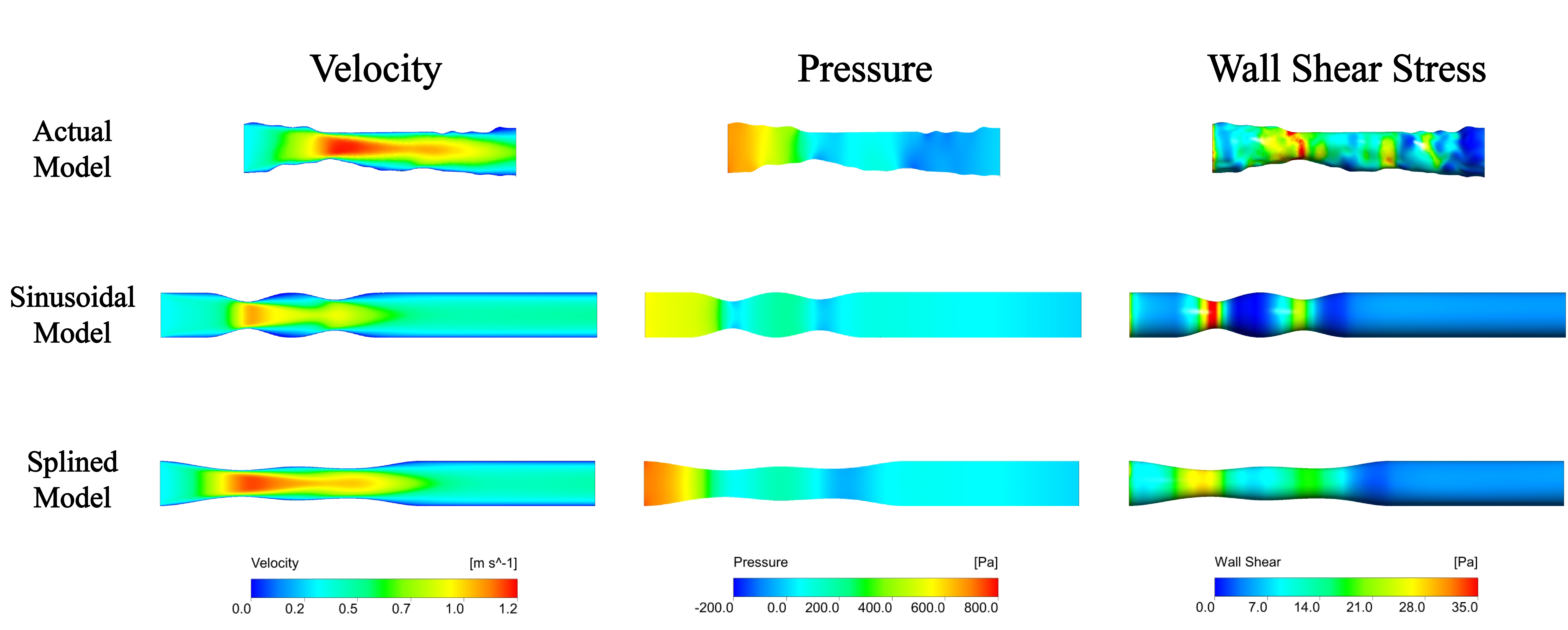}
	\caption{Visualization of flow properties through the actual and simplified double stenosed artery models.}
	\label{fig5}
\end{figure*}

\begin{figure}[h!]
	\centering
	\includegraphics[width=0.5\linewidth]{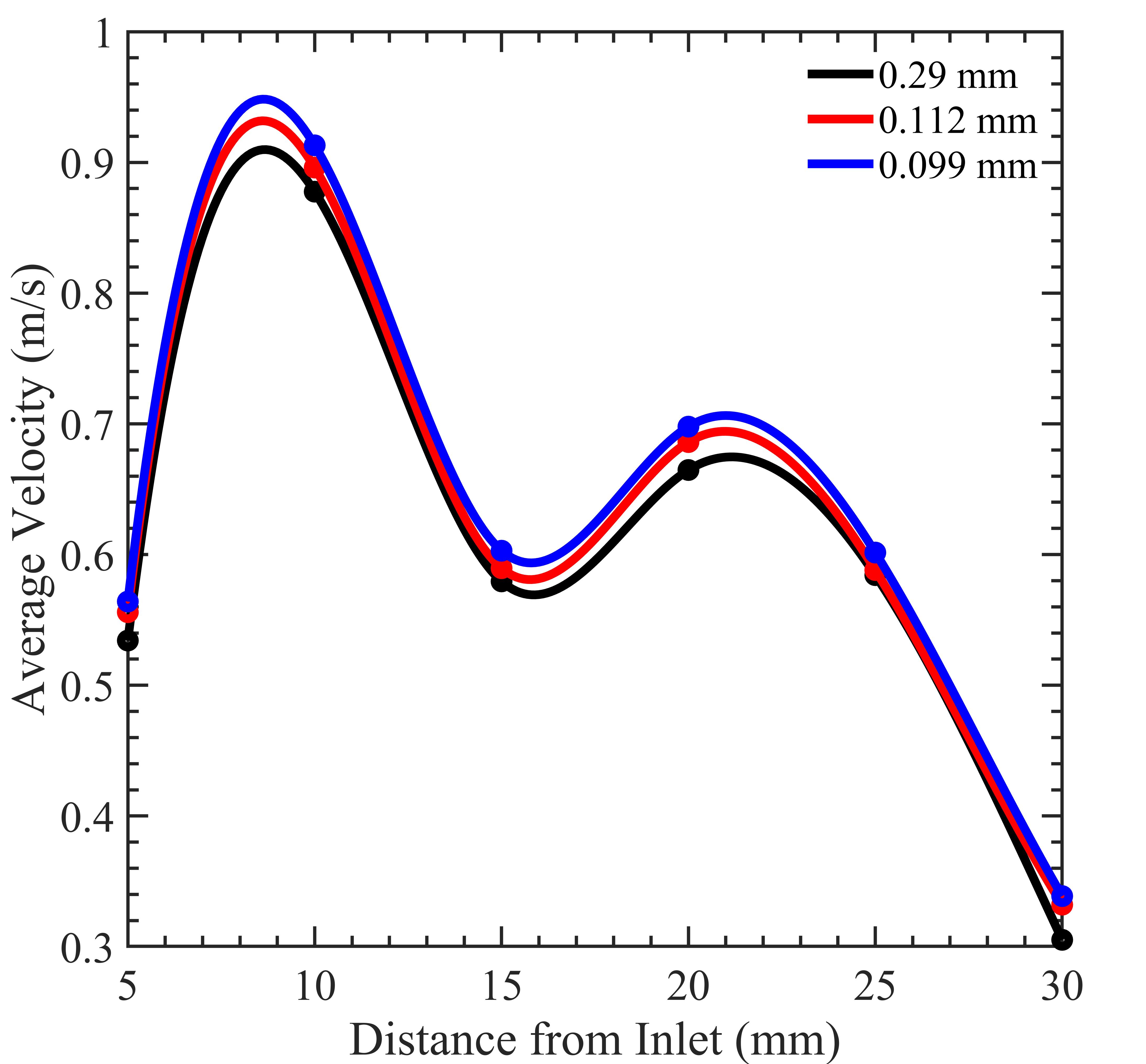}
	\caption{Mesh independence test with various sizes of mesh elements.}
	\label{fig6}
\end{figure}

Fig. \ref{fig5} illustrates the aforementioned flow characteristics visually. The graphic clearly illustrates the influence of naturally produced uneven surfaces on the actual model. Initially, the blood pressure is rather high in all three models. As the blood reaches the first stenosed neck, the pressure gradually drops to or below zero. According to Bernoulli’s principle, the decrease in pressure induces an increase in flow velocity in the stenosed region, as also visible from the figure. This results in a significant rise in $WSS$ at the constriction site. According to the continuity equation, when blood travels further into the first diverging section, the increase in the cross-sectional area results in a reduction in flow velocity. As a consequence, pressure rises and $WSS$ falls. As the blood approaches the next stenosed neck, the velocity increases again but to a lower magnitude due to less constriction there. The pressure decreases once again, along with a slight rise in $WSS$. Further downstream, the velocity attains equilibrium, and both pressure and $WSS$ approach near-zero values. Although not perfect, fig. \ref{fig4} and fig. \ref{fig5} indicate that the splined model would be a better match than the sinusoidal model for the artificial intelligence (Ai) implementation in the next section. Such is the cost of creating a generalized simplified model without being patient-specific. From these discussions, it is also apparent that the progressive change in the cross-sectional diameter of the stenosed aortic vessel gradually affects blood flow characteristics. This enables the generation of a sequential dataset by taking into account the diameters and flow properties at regular intervals along the length of the blood vessel. The mesh independence test, as illustrated in fig. \ref{fig6}, demonstrates that changing the mesh element size has a minor influence on the flow characteristics of blood through the splined double stenosed artery model. Between $0.099\ mm$ and $0.112\ mm$, the variation in simulation results is significantly less. As such, it is more reasonable to adopt a mesh element size of roughly $0.112\ mm$ throughout this study to achieve high CFD simulation accuracy without being too computationally expensive.

\section{Deep Learning Models}
A recurrent neural network (RNN) is a special kind of Ai network with internal memory that enables it to comprehend sequential data. However, the basic RNN is afflicted by a phenomenon known as vanishing gradient \cite{may4}. Long short-term memory (LSTM) and gated recurrent units (GRU) are special kinds of RNN networks developed to mitigate the problem. Their capacity to retain crucial details from the preceding step, such as aortic diameter, would allow them to effectively forecast occurrences in the following step, such as wall shear stress ($WSS$), average velocity ($V_{avg}$) of blood, and pressure. Three techniques are employed in this study to predict these flow properties: Gated Recurrent Unit (GRU), Long short-term memory (LSTM), and Neural Network (NN) models. All three models used inlet velocity and percentage lumen openings at eleven locations along the $50\ mm$ long blood artery at regular $5\ mm$ intervals to predict the blood flow characteristics at those positions. This section highlights each of the three deep learning models.

\subsection{Gated Recurrent Unit model}
Gated Recurrent Unit (GRU) \cite{may2} is a special kind of recurrent neural network that consists of an update gate and a reset gate. GRU's update gate determines how much data from previous units must be passed on. The update gate computes $z_{t}$ for time step $t$ using the formula:
\begin{equation}
z_{t} = \sigma(W_{z}.[h_{t-1}, x_{t}])\label{eq7}
\end{equation}

where $z_{t}$ is update gate output at the current timestamp, $W_{z}$ is weight matrix at update gate, $h_{t-1}$ information from previous units, and $x_{t}$ is input at the current unit.

The model used the reset gate to determine how much information from previous units should be erased. This is calculated using the following formula:
\begin{equation}
  r_{t} = \sigma(W_{r}.[h_{t-1}, x_{t}]) \label{eq8}  
\end{equation}

where $r_{t}$ is reset gate output at current timestamp, $W_{r}$ is weight matrix at reset gate, $h_{t-1}$ information from previous units, and $x_{t}$ is input at the current unit. The relevant data from earlier units were stored in the current memory content using this formula:
\begin{equation}
    \tilde{h}_{t} = tanh (W.[r_{t}*h_{t-1}, x_{t}]) \label{eq9}
\end{equation}

where $h_{t}$ is current memory content, $W$ is weight at current unit, $r_{t}$ is reset gate output at current timestamp, $h_{t-1}$ is information from previous units, and $x_{t}$ is input at the current unit.

Final memory at the current unit was a vector that stored and conveyed the current unit's final information to the next layer. This was computed using the following formula:
\begin{equation}
    h_{t} = (1-z_{t})*h_{t-1} + z_{t}\tilde{h}_{t} \label{eq10}
\end{equation}

where $h_{t}$ is final memory at the current unit, $z_{t}$ is update gate output at current timestamp, $h_{t-1}$ is information from previous units, and $h_{t}$ is current memory content.

\subsection{Long short-term memory model}
Another sort of RNN is the Long short-term memory (LSTM) \cite{may1}. In contrast to the GRU, the LSTM contains three gates: the forget gate, the update gate, and the output gate. The LSTM gates' formulae are as follows:

\begin{equation}
i_{t} = \sigma(W_{i}[h_{t-1}, x_{t}] + b_{i})\label{eq11}
\end{equation}

\begin{equation}
f_{t} = \sigma(W_{f}[h_{t-1}, x_{t}] + b_{f})\label{eq12}
\end{equation}

\begin{equation}
o_{t} = \sigma(W_{o}[h_{t-1}, x_{t}] + b_{o})\label{eq13}
\end{equation}
where $i_{t}$ represents input gate, $f_{t}$ represents forget gate, $o_{t}$ represents output gate, $\sigma$ represents sigmoid function, $W_{x}$ represents weight of the respective gate($x$) neurons, $h_{t-1}$ represents output of previous LSTM block at timestamp $t-1$, $x_{t}$ represents input at current timestamp and $b_{x}$ represents biases for the respective gates($x$).

\begin{figure}[h!]
	\centering
	\includegraphics[width=0.6\linewidth]{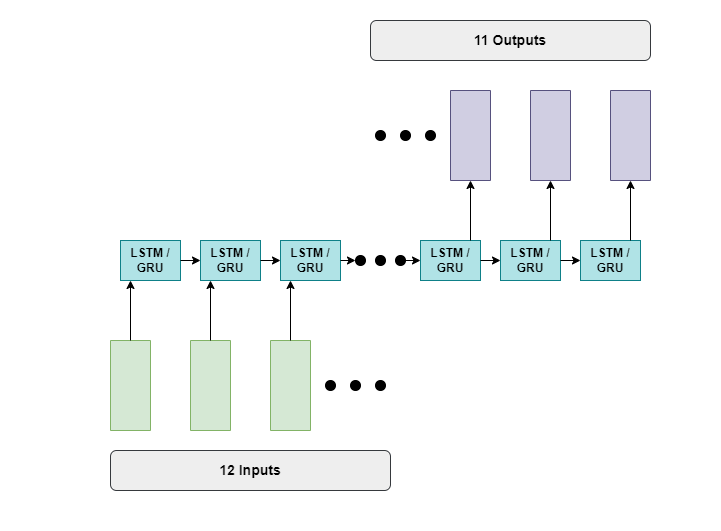}
	\caption{Illustration of LSTM / GRU Ai models in many-to-many configuration.}
	\label{fig7}
\end{figure}

Both GRU and LSTM models utilized the many-to-many combination, with an 11 node dense layer as the output, to predict the $V_{avg}$, $WSS$, or blood pressure at the eleven positions, taking the aforementioned inputs. These models are depicted in fig \ref{fig7}. The hyperparameters of these models were varied in order to maximize the prediction accuracies across all three flow properties. 

\subsection{Neural Network model}
\begin{figure}[h!]
	\centering
	\includegraphics[width=0.6\linewidth]{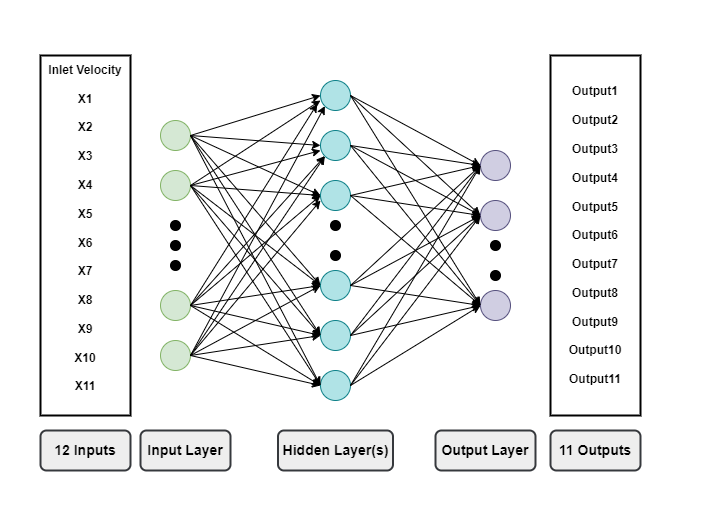}
	\caption{Illustration of neural network (NN) Ai model.}
	\label{fig8}
\end{figure}

The other aspect of the present study is the neural network (NN) as shown in fig. \ref{fig8}. It is defined as a collection of algorithms that are capable of correctly recognizing the underlying connections between a set of data via a method that replicates the way the human brain works. They are not limited to sequential data and are composed of nodes with assigned weights. Through the forward and backward propagation processes using labeled data, the network is able to fine-tune the weights to make accurate predictions. It also, applied the same inputs as the previous two models to predict the same flow features. The hyperparameters were also varied for this model to improve its accuracy.

\section{Dataset}
Since the present study involves a newly proposed doubled stenosed artery model, a dataset of CFD simulation results relating to it is unavailable. As such, a custom dataset containing $180$ data points was constructed. To create the dataset, several configurations of fractions of lumen opening at each stenosed neck, gaps between them, and inlet velocities, as previously mentioned in the simulation setup section, were used. In particular, stenosis severity levels of $25\%$, $50\%$, and $75\%$ were applied at individual necks, with $10\ mm$, $15\ mm$, $20\ mm$, and $25\ mm$ spacing between them. $90\%$ of the total data were utilized for training, while the remaining $10\%$ was used for validation. A different test set containing $18$ datapoints was also constructed using configurations that were absent in the main dataset, such as different inlet velocities and stenosis severity levels of $70\%$, $60\%$, $40\%$, $30\%$, etc. Due to the lack of additional IVUS images with similar double stenosis conditions, the blood vessel employed is a generalized form of a patient-specific actual model. Overcoming such hurdles, as well as including CFD simulations of curved vessels in the future could make the dataset even more beneficial.

\section{Hyperparameters}
In order to improve the accuracy of these three models, several hyperparameters were tuned. Table \ref{tab1} summarizes the ranges of the parameters varied. Firstly, different units for the LSTM / GRU models and different numbers of hidden layers (HL), containing 12 nodes in each, for the NN model were tested to determine the conditions that performed well for predicting all three flow properties. Subsequently, the number of epochs and the learning rates were optimized to maximize the accuracies of the deep learning models. Other parameters such as the loss function, activation functions, and optimizer were kept uniform in all models to provide a fair comparison. Initially, min-max normalization was used to ensure that all of the data in the dataset was within the range of $0$ to $1$.

To compute the loss, each model utilized the Mean squared error (MSE) function which is represented by the formula:
\begin{equation}
    L(y, \hat{y}) = \frac{1}{N} \sum_{i=0}^{N}(y - {\hat{y}}_i)^2 \label{eq14}
\end{equation}

where $\hat{y}$ is the predicted value, $N$ is the number of data points, and $y$ is the observed value. MSE in particular can penalize large errors more than smaller ones, making it a good choice for achieving multiple accurate predictions. Each dense layer utilized the sigmoid activation function, which produces a probabilistic output that exists exclusively between $0$ and $1$, following the equation:

\begin{equation}
    \phi(z) = \frac{1}{1+e^{-z}} \label{eq15}
\end{equation}
For its capability of achieving excellent results quickly and effectively, the adam \cite{may3} optimizer was implemented in all of the models.

\begin{table}[ht]
\centering
\caption{ Hyperparameters adapted in different models.}
\label{tab1}
\begin{tabular}{c c c c}
\hline
\textbf{}
&\textbf{GRU} 
&\textbf{LSTM} 
&\textbf{Neural Network}
\\ 
\hline
\hline\\
& units: 12 & units: 12 & HL: 1\\\cline{2-4}\\
Units / Hidden Layer(HL) & units: 48 & units: 48 & HL: 2\\\cline{2-4}\\
& units: 84 & units: 84 & HL: 3\\\cline{2-4}\\
&units: 120 & units: 120 & HL: 4\\ \hline\\

& 50000 & 50000 & 50000\\\cline{2-4}\\
Epochs & 100000 & 100000 & 100000\\\cline{2-4}\\
& 150000 & 150000 & 150000\\\cline{2-4}\\
&200000 & 200000 & 200000\\ \hline\\

& 0.01 & 0.01 & 0.01\\\cline{2-4}\\
Learning Rate & 0.001 & 0.001 & 0.001\\\cline{2-4}\\
& 0.0001 & 0.0001 & 0.0001\\\cline{2-4}\\
&0.00001 & 0.0001 & 0.00001\\ \hline\\
\end{tabular}
\end{table}
 
\begin{table*}[pt!]
\centering
\caption{ Accuracies achieved by different models by varying the number of units for the GRU and LSTM model and changing the number of hidden layers for the Neural Network model.}
\label{tab2}
\setlength{\tabcolsep}{6pt}
\begin{tabular}{c c c c c c c c c c}
\\ 
\hline\\
\textbf{Model} & \textbf{Units/HL} & \textbf{Epoch} & \textbf{Learning} & & \textbf{Training} & & & \textbf{Testing} & \\ \cline{5-10}
  &  &  & \textbf{Rate} & \textbf{$V_{avg}$} & \textbf{$WSS$} & \textbf{$Pressure$} & \textbf{$V_{avg}$} & \textbf{$WSS$} & \textbf{$Pressure$}\\ \hline \hline
  
   & 12 &  &  & 0.8500 & 0.8778 & 0.8278 & 0.7778 & 0.7222 & 0.8889\\ \cline{2-2} \cline{5-10}
  LSTM & 48 &  &  & 0.8611 & 0.9833 & 0.9722 & 0.8333 & 0.7778 & 0.7778\\ \cline{2-2} \cline{5-10}
   & 84 &  &  & 0.8778 & 0.9778 & 0.9889 & 0.8889 & 0.8889 & 0.8889\\ \cline{2-2} \cline{5-10}
   & 120 & &  & 0.9111 & 0.9722 & 0.9444 & 0.9444 & 0.7222 & 0.8333\\ \cline{1-2} \cline{5-10}
   
   & 12 &  &  & 0.7944 & 0.7667 & 0.9500 & 0.8333 & 0.7222 & 0.9444\\ \cline{2-2} \cline{5-10}
  GRU & 48 &  &  & 0.9000 & 0.9722 & 0.9722 & 0.9444 & 0.7778 & 0.9444\\ \cline{2-2} \cline{5-10}
   & 84 & 100000 & 0.0001 & 0.8833 & 0.9556 & 0.9556 & 0.7778 & 0.8333 & 0.9444 \\ \cline{2-2} \cline{5-10}
   & 120 &  &  & 0.8389 & 0.9667 & 0.9611 & 0.9444 & 0.8889 & 0.9999\\ \cline{1-2} \cline{5-10}

  & 1 &  &  & 0.8444 & 0.9500 & 0.7167 & 0.8333 & 0.8333 & 0.7222\\ \cline{2-2} \cline{5-10}
  NN & 2 &  &  & 0.8222 & 0.9722 & 0.9389 & 0.8333 & 0.7778 & 0.9999\\ \cline{2-2} \cline{5-10}
   & 3 &  &  & 0.8222 & 0.9667 & 0.8778 & 0.7778 & 0.7222 & 0.9999\\ \cline{2-2} \cline{5-10}
   & 4 &  &  & 0.8167 & 0.8278 & 0.9333 & 0.8889 & 0.5556 & 0.9999\\ \hline
\end{tabular}
\end{table*}

\begin{table*}[pt!]
\centering
\caption{ Accuracies achieved by different models by varying the number of epochs.}
\label{tab3}
\setlength{\tabcolsep}{6pt}
\begin{tabular}{c c c c c c c c c c}
\\ 
\hline\\
\textbf{Model} & \textbf{Units/HL} & \textbf{Epoch} & \textbf{Learning} & & \textbf{Training} & & & \textbf{Testing} & \\ \cline{5-10}
  &  &  & \textbf{Rate} & \textbf{$V_{avg}$} & \textbf{$WSS$} & \textbf{$Pressure$} & \textbf{$V_{avg}$} & \textbf{$WSS$} & \textbf{$Pressure$}\\ \hline \hline
  
   &  & 50000 &  & 0.8667 & 0.9389 & 0.7000 & 0.7778 & 0.7222 & 0.5000\\ \cline{3-3} \cline{5-10}
  LSTM & 84 & 100000 &  & 0.8778 & 0.9778 & 0.9889 & 0.8889 & 0.8889 & 0.8889\\ \cline{3-3} \cline{5-10}
   & & 150000 &  & 0.8944 & 0.9778 & 0.9944 & 0.7778 & 0.7222 & 0.9999\\ \cline{3-3} \cline{5-10}
   & & 200000 &  & 0.9222 & 0.9778 & 0.9722 & 0.7778 & 0.7778 & 0.7222\\ \cline{1-3} \cline{5-10}
   
   &  & 50000 &  & 0.8389 & 0.9389 & 0.8222 & 0.8333 & 0.9444 & 0.7778\\ \cline{3-3} \cline{5-10}
  GRU & 48 & 100000 &  & 0.9000 & 0.9722 & 0.9722 & 0.9444 & 0.7778 & 0.9444\\ \cline{3-3} \cline{5-10}
   & & 150000 & 0.0001 & 0.8667 & 0.9556 & 0.9667 & 0.8889 & 0.7778 & 0.9999\\ \cline{3-3} \cline{5-10}
   & & 200000 &  & 0.8944 & 0.9667 & 0.9778 & 0.8889 & 0.8889 & 0.9999\\ \cline{1-3} \cline{5-10}
   
  &  & 50000 &  & 0.8500 & 0.9333 & 0.6444 & 0.8333 & 0.8333 & 0.6667\\ \cline{3-3} \cline{5-10}
  NN & 2 & 100000 &  & 0.8222 & 0.9722 & 0.9389 & 0.8333 & 0.7778 & 0.9999\\ \cline{3-3} \cline{5-10}
   & & 150000 &  & 0.8222 & 0.9389 & 0.9778 & 0.8333 & 0.9444 & 0.9999\\ \cline{3-3} \cline{5-10}
   & & 200000 &  & 0.8556 & 0.9778 & 0.9999 & 0.7778 & 0.8333 & 0.9999\\ \hline
\end{tabular}
\end{table*}

\begin{table*}[pt!]
\centering
\caption{ Accuracies achieved by different models by varying the learning rate.}
\label{tab4}
\setlength{\tabcolsep}{6pt}
\begin{tabular}{c c c c c c c c c c}
\\ 
\hline\\
\textbf{Model} & \textbf{Units/HL} & \textbf{Epoch} & \textbf{Learning} & & \textbf{Training} & & & \textbf{Testing} & \\ \cline{5-10}
  &  &  & \textbf{Rate} & \textbf{$V_{avg}$} & \textbf{$WSS$} & \textbf{$Pressure$} & \textbf{$V_{avg}$} & \textbf{$WSS$} & \textbf{$Pressure$}\\ \hline \hline
  
   &  & & 0.01 & 0.8611 & 0.9944 & 0.9944 & 0.9444 & 0.7778 & 0.8889\\ \cline{4-10}
  LSTM & 84 & 100000 & 0.001 & 0.9500 & 0.9778 & 0.9778 & 0.7778 & 0.8889 & 0.9999\\\cline{4-10}
   & &  & 0.0001 & 0.8778 & 0.9778 & 0.9889 & 0.8889 & 0.8889 & 0.8889\\ \cline{4-10}
   & & & 0.00001 & 0.8833 & 0.9500 & 0.500 & 0.8333 & 0.7222 & 0.6111\\ \hline
   
   &  & & 0.01 & 0.8389 & 0.9722 & 0.9778 & 0.8333 & 0.5556 & 0.9999\\ \cline{4-10}
  GRU & 48 & 100000 & 0.001 & 0.9333 & 0.9833 & 0.9833 & 0.9444 & 0.8333 & 0.9444\\\cline{4-10}
   & &  & 0.0001 & 0.9000 & 0.9722 & 0.9722 & 0.9444 & 0.7778 & 0.9444\\ \cline{4-10}
   & & & 0.00001 & 0.8611 & 0.9111 & 0.7778 & 0.8889 & 0.6667 & 0.8333\\ \hline
   
  &  & & 0.01 & 0.8944 & 0.9833 & 0.3611 & 0.7778 & 0.6111 & 0.2778\\ \cline{4-10}
  NN & 2 & 150000 & 0.001 & 0.8111 & 0.9944 & 0.9999 & 0.8889 & 0.8333 & 0.9999\\\cline{4-10}
   & &  & 0.0001 & 0.8222 & 0.9389 & 0.9778 & 0.8333 & 0.9444 & 0.9999\\ \cline{4-10}
   & & & 0.00001 & 0.7111 & 0.6111 & 0.4222 & 0.6111 & 0.5556 & 0.4444\\ \hline
\end{tabular}
\end{table*}

\section{Results and Discussions}
Training and testing accuracies were used to assess the three suggested Deep Learning approaches. These accuracies indicated the degree to which each model could correctly predict $V_{avg}$, $WSS$, or blood pressure during testing or training. The variance in the accuracies for varying the number of units in the GRU and LSTM models, as well as for different the number of hidden layers in the NN model is represented in Table \ref{tab2}. In these cases, the number of epochs and learning rate were kept constant at $100000$ and $0.0001$ respectively. Owing to a small dataset, the LSTM model with $84$ units overall performed well across all flow properties for both training and test sets, as visible in the table. In contrast, the GRU model did well throughout both sets with only $48$ units. Extending the units beyond these ideal values seems to degrade the efficacy of both models. On the other hand, the NN model shows an overall decreased effectiveness in predicting the flow properties for both sets as compared to the other two models with the same number of epochs and learning rate. Nonetheless, testing with different numbers of hidden layers reveals that the smaller number of data prefers a shallower architecture with $2$ intermediate layers only, having good accuracies on most occasions. Pressure values in the dataset were either very large or very small at certain points along the length of the vessel, as also visible from fig \ref{fig4}. As a consequence, when these numbers were normalized, tiny values were transformed to near-zero values, while bigger values were changed to near-one values. As such, each model only needed to learn either of these two extreme quantities at certain locations, leading to much higher accuracies for both training and test sets.

To further tune the hyperparameters and improve the accuracies of the models, the number of epochs was varied from $50000$ to $200000$ while keeping the learning rate constant at $0.0001$. Based on the previous evaluations, the number of units for LSTM, GRU, and the number of hidden layers for the NN model is set to $84$, $48$, and $2$ respectively. The results of these evaluations are reported in Table \ref{tab3}. It can be seen that for the LSTM model, the highest accuracy for $V_{avg}$ is obtained at 200000 epochs, whereas pressure achieves the greatest accuracy at 150000 epochs. The predicting effectiveness for $WSS$ remains unchanged from $100000$ to $200000$ epochs. However, the average testing accuracies across three properties are much lower above and below $100000$ epochs, indicating the model cannot generalize well in those rangers. The GRU on average also performs very well at $100000$ epochs using both sets of data. The training dataset for the NN model clearly prefers the higher epochs but the peak average test set accuracy at $150000$ epochs indicates non generalizing effect beyond this value.

Finally, the learning rates of each model are tuned, setting the other hyperparameters to their predetermined optimum values. The results of these investigations are shown in Table \ref{tab4}. The table suggests that LSTM performs very well in predicting the flow properties for learning rate in between $0.01$ to $0.0001$, whereas GRU is most accurate for a rate of $0.001$. NN model on the other hand prefers a learning rate between $0.001$ and $0.0001$. In certain cases the larger learning rate overshoots, destabilizes the training process, and fails to reach optimum accuracy, for example, NN for predicting pressure of the flow. It also tends to overfit the data from both LSTM and NN models. Too small learning rate isn’t beneficial either since it also lowers the overall accuracies of all the models across the three flow properties. Although not the most accurate in every instance, a learning rate of $0.001$ that can generalize well is preferred by all three models to reasonably predict each of the flow properties from both datasets.

\begin{figure}[t!]
	\centering
	\includegraphics[width=0.98\linewidth]{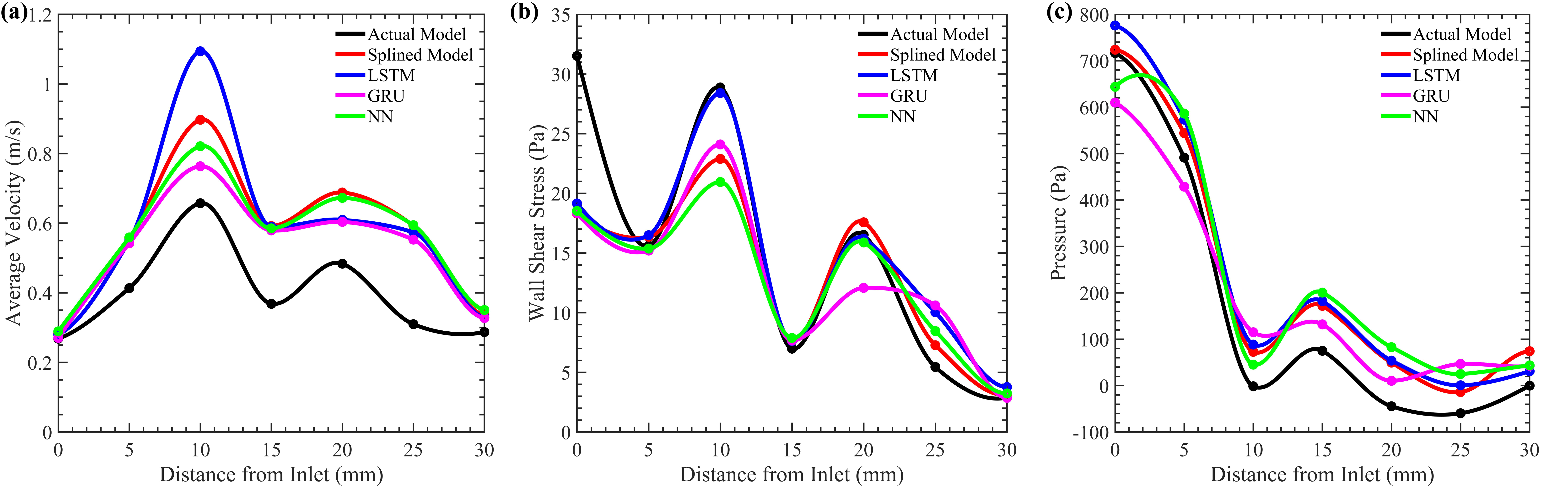}
	\caption{Comparison between predicted flow characteristics with optimized hyperparameters and simulated results of actual and splined artery models.}
	\label{fig9}
\end{figure}

Fig. \ref{fig9} compares the predicted flow characteristics obtained with the optimized hyperparameters to the CFD simulation results obtained with the real and splined models for the identical configuration with $40.25\%$ and $32\%$ stenosis located $10\ mm$ apart. In the case of average velocity, the LSTM model overestimates, while the GRU model underestimates, and therefore fails to effectively estimate the flow pattern. On the other hand, the NN model is far more accurate at forecasting flow velocity patterns and is comparable to the simulation results achieved with the splined model. Again, for $WSS$, the NN models perform well and closely track the simulation outcomes. Interestingly, the prediction imperfections put the LSTM model's $WSS$ pattern prediction closer to the actual model's simulation outcomes without even training on that model. Nevertheless, both LSTM and GRU models fail to predict the $WSS$ pattern entirely. Then again, the LSTM model performs much better at predicting pressure fluctuations, while the GRU model's simplicity prevents it from learning the pressure changes effectively from the small dataset. Unlike the other two properties, the NN model seems to be less capable of comprehending flow pressure variation. Thus, fig. \ref{fig9} further demonstrates that tuning the hyperparameters to generalize the models for predicting all flow properties leads to their overall diminished performance. As such, although more effort is necessary, optimizing the models based on each of the flow characteristics individually would be more beneficial. Additionally, the graphic illustrates the NN model's supremacy in predicting short lengths of sequential data from a small dataset.

\section{Conclusion}
The present study explored the faithfulness of simplified models’ flow characteristics to that of the actual model derived from IVUS imaging. The model with a sinusoidal representation of stenosed geometry, which has been widely used in earlier research, entirely fails to portray the actual model's fluctuations in flow property patterns. Although not fully perfect, owing to non-circular cross-sections of the actual model, the newly proposed splined model stands out as a better representation for the construction of a database with various percentage stenoses with varying gaps between them, for implementation of artificial intelligence. The sequential nature of the input and subsequently the flow properties opened up opportunities for specialized RNNs to be implemented. As it turns out, the short lengths of the vessel and small dataset prefer a simpler, less sophisticated conventional neural network model with shallow architecture for efficiently predicting most flow parameters, such as average velocity and wall shear stress. On the other hand, the considerable variation in pressure along the short length of the vessel favors the computationally expensive LSTM model with a large number of units. The simpler GRU model, although generalized well in terms of over accuracies across both training and test sets, fails to generate better predictions than the other two Ai models for any individual double stenosed artery. This highlights the fact that instead of aiming to achieve overall good performance across all outputs with a single set of hyperparameters, each property needs to be addressed and models optimized individually.


\section*{Acknowledgment}

The authors would like to express their gratitude to Dr. Md. Habibur Rahman, Associate Professor and Senior Consultant Cardiologist at National Heart Foundation Hospital \& Research Institute, for providing medical expertise and IVUS imaging for this study.







\end{document}